\documentclass[11pt,twoside]{article}
\usepackage{CAGN2019}
\usepackage{graphicx}

\usepackage[T1]{fontenc} 

\usepackage{latexsym}
\usepackage{verbatim}

\usepackage{ifpdf}  
\ifpdf  
  \DeclareGraphicsExtensions{.pdf,.png,.jpg}  
\else  
  \DeclareGraphicsExtensions{.eps}  
\fi 

\newcommand{\elecd}{$n_{\rm e}$} 
\newcommand{\hii}{H\thinspace{\sc ii}} 
\newcommand{\elect}{$T_{\rm e}$} 
 
\newcommand{\foiii}{[O\thinspace{\sc iii}]}

\newcommand{\fnii}{[N\thinspace{\sc ii}]} 
\newcommand{\fniii}{[N\thinspace{\sc iii}]} 
\newcommand{\oii}{O\thinspace{\sc ii}} 
\newcommand{\cii}{C\thinspace{\sc ii}} 
\newcommand{\hei}{He\thinspace{\sc i}}

\begin{document}

\vskip 1.0cm
\markboth{C.~Esteban et al.}{Galactic abundance gradients}
\pagestyle{myheadings}
%
%
\vspace*{0.5cm}
\parindent 0pt{Invited Review}


\vspace*{0.5cm}
\title{Galactic abundance gradients from deep spectra of {\hii} regions}

\author{C.~Esteban$^{1, 2}$, J.~Garc\'ia-Rojas$^{1, 2}$, K.~Z.~Arellano-C\'ordova $^{1, 2, 3}$ and J.E.~M\'endez-Delgado$^{1, 2}$}
\affil{$^1$Instituto de Astrof\'isica de Canarias, E-38200 La Laguna, Spain\\
$^2$Departamento de Astrof\'isica, Universidad de La Laguna, E-38206, La Laguna, Spain\\
$^3$Instituto Nacional de Astrof\'isica, \'Optica y Electr\'onica, Apdo. Postal 51 y 216, Puebla, Mexico}

\begin{abstract}
We present some results of an ongoing project devoted to reassess the radial abundance gradients in the disc of the Milky Way based on deep spectroscopy of {\hii} regions.
The data have been taken with large aperture telescopes, mainly with the GTC and VLT. The sample contains about 35 objects  located at Galactocentric distances from 5 to 17 kpc. 
We determine the electron temperature for all nebulae, allowing the precise calculation of chemical abundances. In this paper, we present and discuss results mostly 
concerning the radial gradients of O, N and C. 


\end{abstract}

\section{Introduction}

Since decades ago, we know that chemical abundances in spiral galaxies show well defined radial gradients \cite[e.g.][]{searle71, pageledmunds81}. These gradients 
trace the spatial distribution and temporal evolution of star formation history and the effects of gas flows and other processes over the chemical composition of the 
galaxies. {\hii} regions can be used as probes of the present-day composition of the interstellar medium in the zone of 
the hosting galaxy where they are located. The emission-line spectrum of {\hii} regions can be used to determine the gas-phase abundance of several elements, 
especially of O -- the proxy of metallicity in the analysis of ionized nebula -- but also of He, C, N, Ne, S, Cl, Ar and Fe. 

In the late seventies, several works investigated the radial abundance gradients in the Milky Way based on optical spectroscopy of a small number of {\hii} regions
\cite[e.g.][]{peimbertetal78, hawley78, talentdufour79}. Until recently, the number of Galactic {\hii} regions with direct determination of the electron temperature, {\elect}, 
from optical spectra has been rather limited. In fact, the classic work on the subject by \cite{shaveretal83}, who observed a larger sample of {\hii} regions in a 
wider range of Galactocentric distances, used optical spectra to measure the collisionally excited lines (hereinafter CEL) of heavy-element ions but radio recombination 
line measurements to derive  the {\elect} of the nebulae. Optical and radio observations were not co-spatial and the aperture sizes of both kinds of data were very different. 
\cite{deharvengetal00} also combined non-co-spatial optical spectra and radio observations to derive {\elect} for most of the 34 {\hii} regions of their sample. The studies 
of the radial abundance gradients based on far infrared, FIR, observations \cite[e.g.][]{rudolphetal06} combine measurements of CELs from FIR 
spectra and {\elect} determined from radio observations with also different apertures. 

As we have said above, the O/H ratio is the proxy of metallicity in nebular abundance studies. In normal {\hii} regions, O abundance can be derived
simply adding the O$^+$ and O$^{2+}$ abundances, that can be obtained from the intensity of bright optical CELs. The paucity of abundance determinations based on  
co-spatial direct determinations of {\elect} for {\hii} regions, especially at both extremes of the Galactic disc -- central zones and 
anticentre -- has been an enduring problem in the exploration of the true shape of the Galactic O gradient. Those distant nebulae are usually faint and heavily reddened 
and the direct determinations of {\elect} -- essential for determining reliable abundances -- is very difficult  \cite[see][]{peimbertetal78, estebanetal05}. 

\section{The sample and methodology}

The aim of our ongoing project is to obtain very deep spectroscopy of a sizable sample of {\hii} regions covering the largest possible fraction of the Galactic disc. By now, the observed sample covers 35 objects located at Galactocentric distances, $R_{\rm G}$, between 5 and 17 kpc. The preliminary results concerning the O and N gradients have been published in \cite{estebanetal17} and \cite{estebangarciarojas18}. Thirteen objects of the sample are located beyond the isophotal radius of the Milky Way \cite[$R_{25}$ = 11.5 kpc,][]{devaucouleurspence78} and have been selected to explore the shape of the radial abundance gradients at the Galactic anticentre. Almost all the spectra have been taken with high or intermediate-resolution spectrographs attached to 8 - 10 m telescopes (GTC and VLT). In all the {\hii} regions the {\elect}-sensitive auroral {\foiii} 4363 \AA\ and/or {\fnii} 5755 \AA\ lines have been measured, allowing us to obtain a precise determination of {\elect} and the ionic abundances and avoiding uncertainties due to the combination of non-co-spatial data. 

We have been especially careful in selecting appropriate values of $R_{\rm G}$ for the objects. For each {\hii} region, we have assumed the mean values of the kinematic and stellar distances given in different published references \cite[see][for details]{estebanetal17, estebangarciarojas18}. We have associated an uncertainty for each distance, which corresponds to the standard deviation of the values used for the mean. In contrast to what is customary in many previous works, we include the errors in $R_{\rm G}$ when calculating the linear fits of the gradients. We have assumed the Sun located at $R_{\rm G}$ = 8.0 kpc \citep{reid93}. 

We have derived the electron density, {\elecd},  and {\elect} using the density and temperature-sensitive emission line ratios of the CELs observed in each spectrum. Using these lines, we can also derive ionic and total abundances of several elements as He, C, N, O, Ne, S, Cl, Ar and Fe.  We use the same methodology and atomic dataset to derive physical conditions and abundances for all the objects. 

\begin{figure}
\center
\includegraphics[scale=0.5]{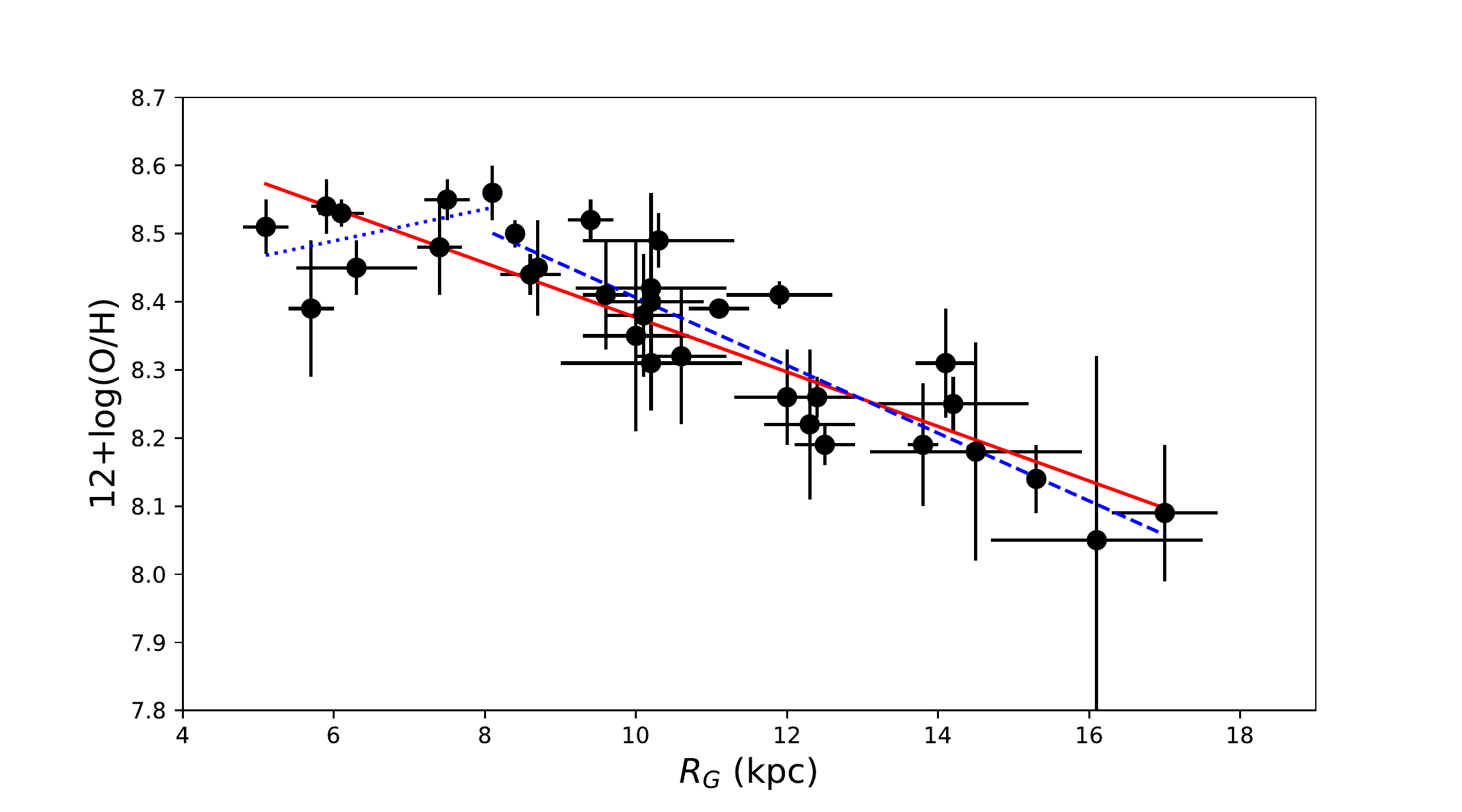}
\caption{\label{Ograd}Radial distribution of the O abundance -- in units of 12+log(O/H) -- as a function of the Galactocentric distance, $R_{\rm G}$, for our sample of Galactic {\hii} regions. The solid red line represents the least-squares fit to all objects. The dashed blue line corresponds to the least-squares fit to the {\hii} regions located at  $R_{\rm G}$ $>$ 8 kpc and the dotted blue line to those with $R_{\rm G}$ $<$ 8 kpc.}
\end{figure}

\section{The O/H gradient}

In Fig.~\ref{Ograd} we present the spatial distribution of the O/H ratio for the {\hii} regions of our sample 
\citep[a first version of this graph was published in][]{estebangarciarojas18}. The least-squares linear fit to the $R_{\rm G}$ and the O abundances, gives the following radial gradient :

\begin{equation} \label{eq1} 12 + \log(\mathrm{O/H}) = 8.80(\pm 0.09) - 0.041(\pm 0.006) R_\mathrm{G}; \end{equation}

\noindent valid for $R_{\rm G}$ from 5.1 to 17.0 kpc. The slope of the gradient is consistent with other previous determinations, as those obtained by \cite{deharvengetal00} or \cite{estebanetal05}. As it was reported by \cite{estebanetal17}, the slope of the radial O abundance gradient does not change for objects located beyond $R_{25}$. This fact demonstrates the absence of flattening of the O gradient in the outer Milky Way, at least up to $R_{\rm G}$ $\sim$ 17 kpc, contrary to what was claimed in previous works \cite[e. g.][]{fichsilkey91, vilchezesteban96, macieletal06}. In Fig.~\ref{Ograd}, we can note that the radial distribution of the O/H ratio seems to change its slope for objects located at $R_{\rm G}$ $<$ 8 kpc. In fact, this zone shows a drop or flattening of the O gradient. We have made a double linear fit of the spatial distribution of the O abundances. Firstly, we made a least-squares linear fit to the $R_{\rm G}$ and the O/H ratios for objects with $R_{\rm G}$ $>$ 8 kpc (represented by a dashed blue line in Fig.~\ref{Ograd}):

\begin{equation} \label{eq2} 12 + \log(\mathrm{O/H}) = 8.90(\pm 0.11) - 0.050(\pm 0.010) R_\mathrm{G}; \end{equation}

\noindent which is somewhat stepper than the fit we obtain for the whole sample but still consistent within the errors. The least-squares linear fit for the objects located in the inner part of the Galactic disc, with $R_{\rm G}$ $<$ 8 kpc, is (represented by a dotted blue line in Fig.~\ref{Ograd}): 

\begin{equation} \label{eq3} 12 + \log(\mathrm{O/H}) = 8.35(\pm 0.13) + 0.023(\pm 0.019) R_\mathrm{G}. \end{equation}

The presence of a drop or flattening of the O/H ratio in the inner zones of the Galactic disc may have important implications for chemical evolution models. 
There are previous works that have found indications of a change on the abundance distribution of O, Fe and $\alpha$-elements in Cepheids in the inner Galactic disc 
\citep{martinetal15, andrievskyetal16}. Another evidence can be found in metallicity gradients derived from SDSS-III/APOGEE observations of red giants by 
\cite{haydenetal14}. These authors find an apparent flattening at $R_{\rm G}$ $<$ 6 kpc, especially evident for low-[$\alpha$/$Z$] stars. The flattening found with 
Cepheids or red giants begins at somewhat smaller distances (at $R_{\rm G}$ $\sim$ 5-6 kpc) than suggested by our {\hii} region data, but the results seem to be 
qualitatively consistent considering the uncertainties. \cite{andrievskyetal16} proposed that this change of slope could be due to a decrease or quenching of the star 
formation rate produced by gas flows towards the Galactic Centre induced by the presence of the Galactic bar. On the other hand, \cite{haywoodetal16} and 
\cite{khoperskovetal18} propose that a quenching of the star formation may be produced by the increasing of turbulence in the gas due to the stellar bar. 
A third evidence of a drop of O/H ratios at inner parts of the Milky Way comes from abundance studies of planetary nebulae, PNe, in the Galactic bulge. 
Using mid-infrared {\it Spitzer} spectra, \cite{gutenkunstetal2008} found that the mean O abundance of these PNe -- located at $R_{\rm G}$ between 0 and 3 kpc -- do not follow the trend of the disc and is about or even slightly lower than solar. The effect of the Galactic bar has also been advocated as the possible origin of this behavior  \citep{pottaschbernardsalas15}.

Inner drops in the radial O abundance distribution have been reported in several spiral galaxies \cite[e.g.][]{belleyroy92, rosalesortegaetal11, sanchezmenguianoetal18}. In all the cases, these features have been obtained from abundance analysis based on strong-line methods and not on direct determinations of {\elect} of the {\hii} regions. \cite{sanchezmenguianoetal18} have found that about 35\% of a sample of about 100 galaxies show an inner drop located about half of the effective radius, $R_{\rm e}$, of the galaxy. Considering that $R_{\rm e}$ is between 4-5 kpc in the Milky Way \citep{devaucouleurspence78}, the position of our change of slope is located at a considerably larger distance than expected if the behavior found by \cite{sanchezmenguianoetal18} is extrapolated to our Galaxy.

\begin{figure} 
\begin{center}
\includegraphics[angle=0,width=6.0cm]{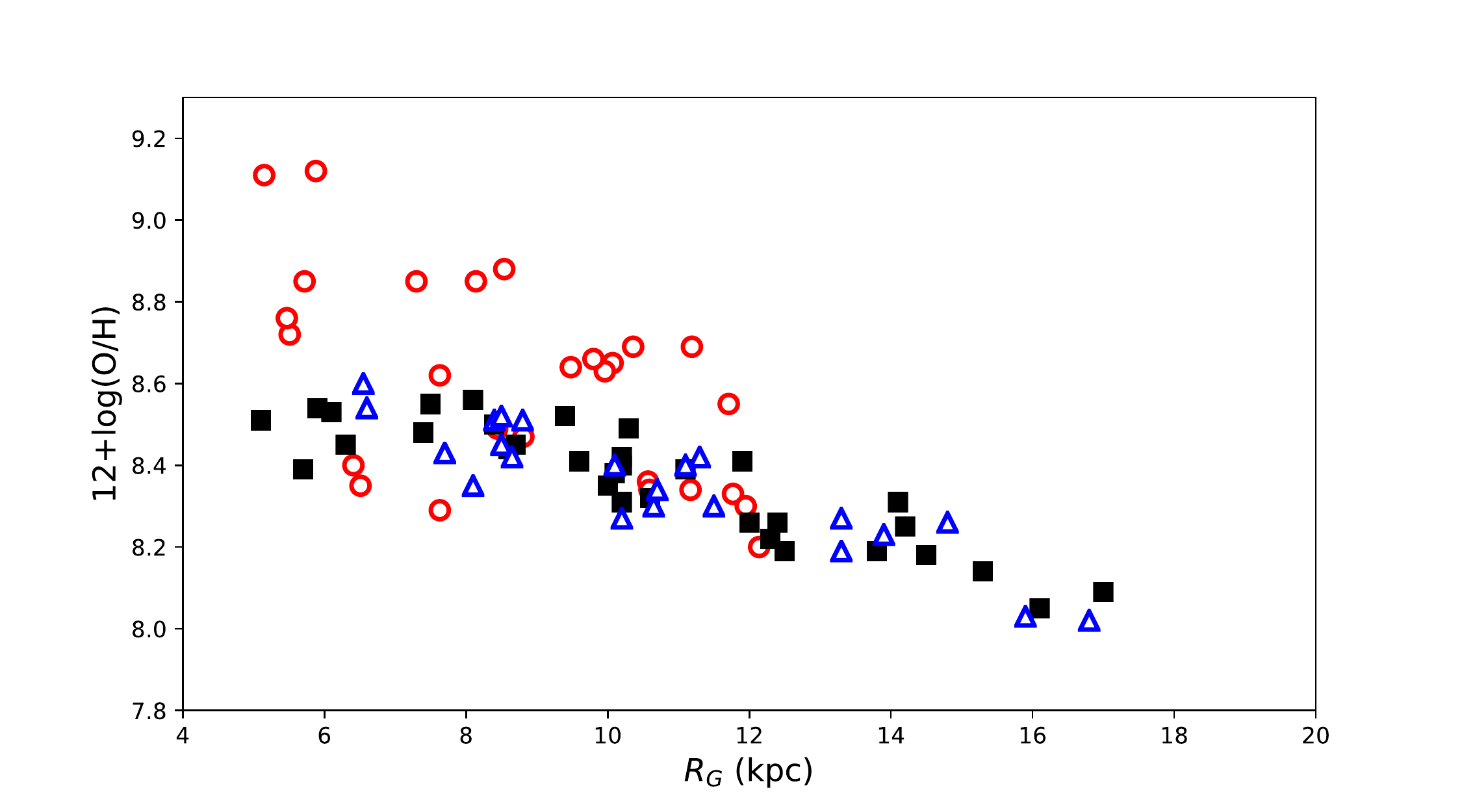}
\hspace*{0.cm}
\includegraphics[width=6.0cm]{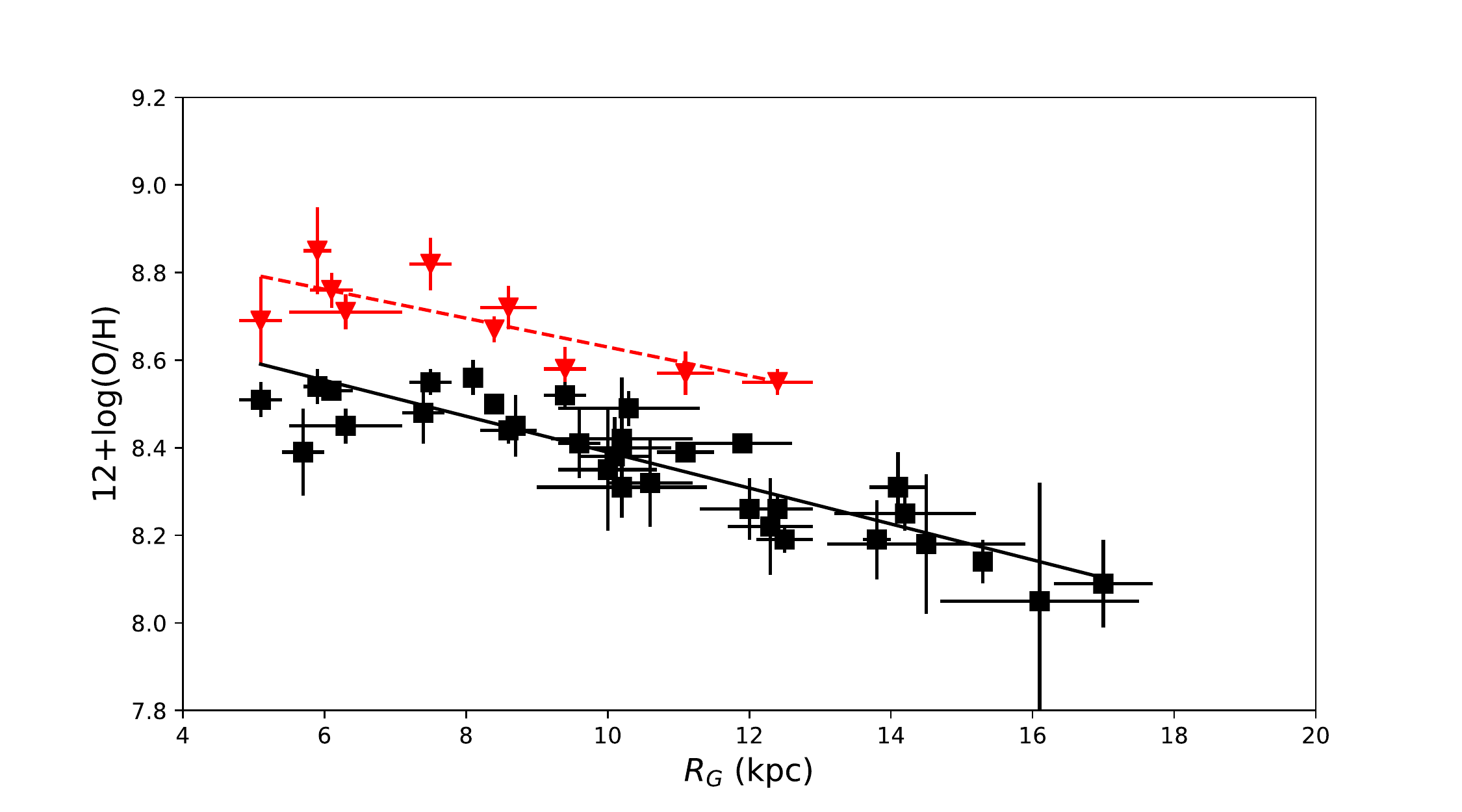}
\caption{ {\it Left:} Radial distribution of the O abundance as a function of $R_{\rm G}$ including our sample of Galactic {\hii} regions (black squares); the blue empty triangles represent data from \cite{deharvengetal00} and the red empty circles include data from  \cite{shaveretal83, peimbertetal78, hawley78} and \cite{talentdufour79}. 
{\it Right:}  Comparison of the radial gradient of O/H obtained from CELs (black squares and continuous line) and RLs (red triangles and dashed line).}
\label{Ograd_comps}
\end{center}
\end{figure}

The mean difference of the O abundance of the {\hii} regions represented in Fig.~\ref{Ograd} and the abundance given by Eq.~\ref{eq1} at their corresponding distance is $\pm$0.05 dex, of the order of the typical abundance uncertainties. That low dispersion indicates that O is well mixed in the interstellar gas along the observed section of the Galactic disc. This result contrast dramatically, for example, with the large scatter shown in the data by \cite{shaveretal83} (Fig.~\ref{Ograd_comps}). 

In Fig.~\ref{Ograd_comps} we also show the comparison between the radial O gradient obtained for our sample objects when using CELs or RLs of O$^{2+}$. The gradient obtained from RLs includes the objects studied in \cite{estebanetal05, estebanetal13} and Sh~2-100, which data are described in \cite{estebanetal17}. 
As we can see, both gradients are almost parallel, in fact the slope obtained from RLs is $-$0.033 $\pm$ 0.009 dex kpc$^{-1}$. The mean offset between both gradient 
lines is of the order of 0.2 dex, the typical value of the abundance discrepancy factor for {\hii} regions \cite[see paper by Garc\'ia-Rojas et al. in these proceedings or][]{garciarojasesteban07}. 

\section{The N/H and N/O gradients determined without ICF}

\begin{figure} 
\begin{center}
\includegraphics[angle=0,width=6.0cm]{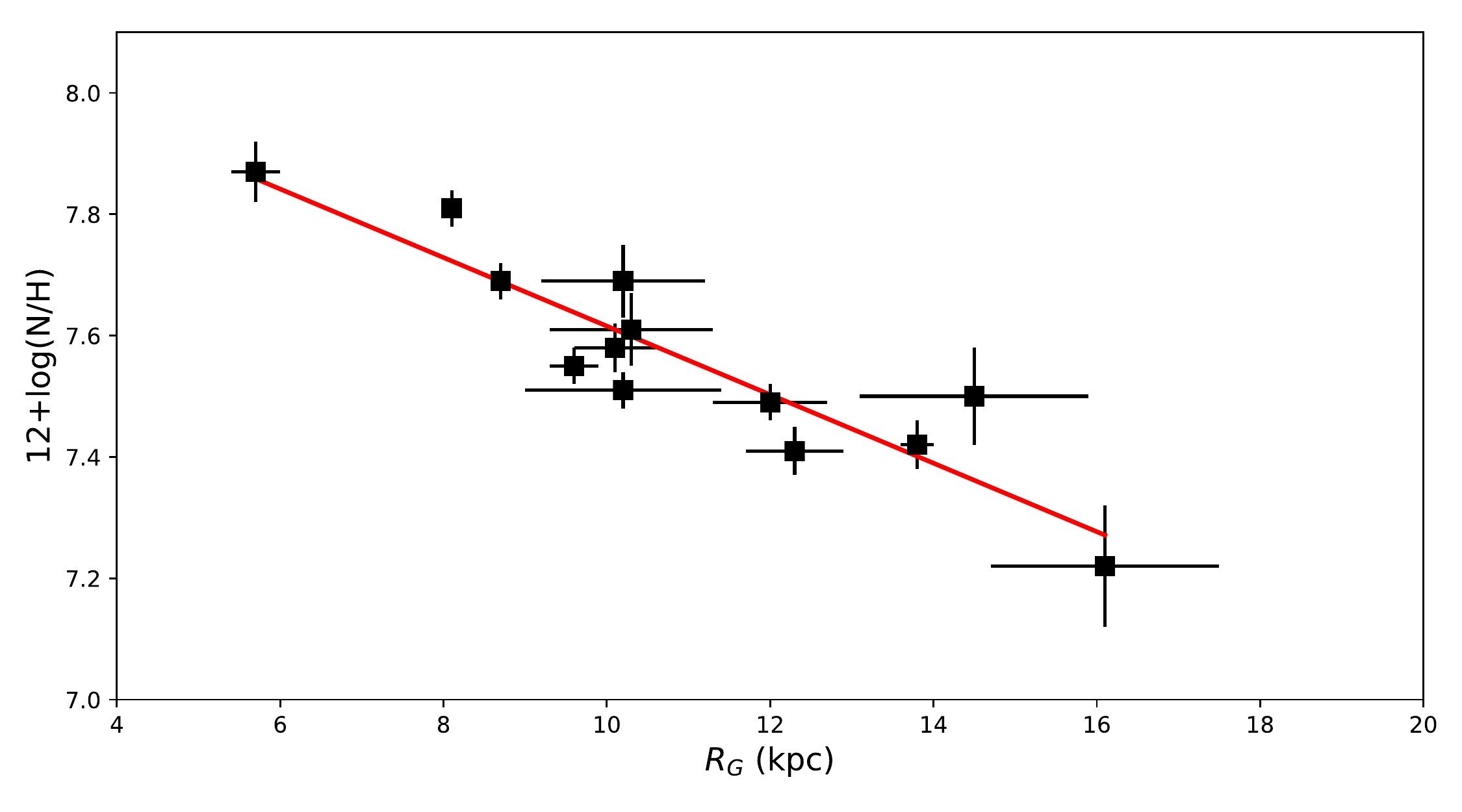}
\hspace*{0.cm}
\includegraphics[width=6.0cm]{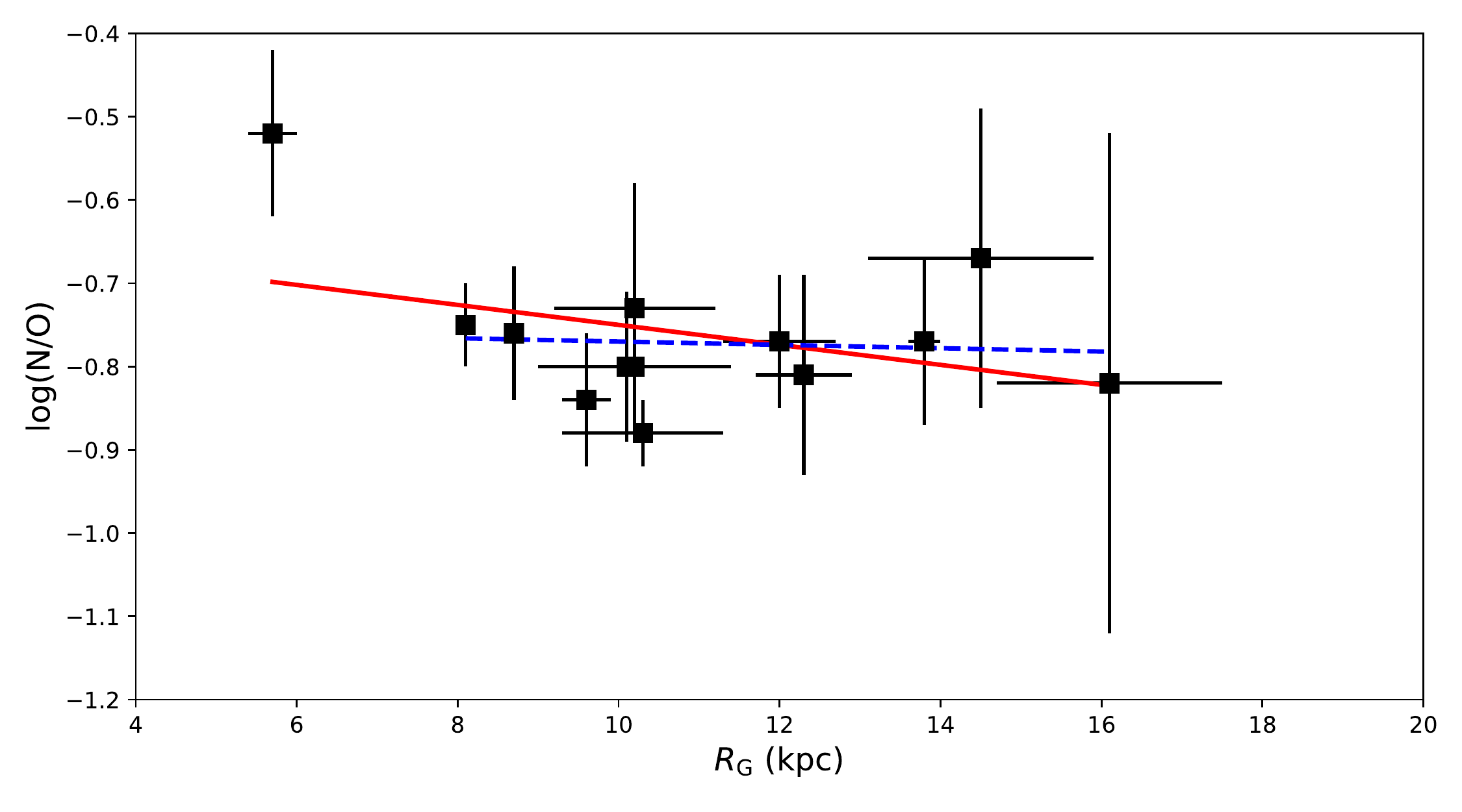}
\caption{ {\it Left:} Radial distribution of the N abundance as a function of $R_{\rm G}$ for our low-ionization sample of Galactic {\hii} regions. In these objects  
N/H $\approx$ N$^+$/H$^+$. The solid red line represents the least-squares fit to all objects. 
{\it Right:}  Radial distribution of log(N/O) as a function of $R_{\rm G}$, for the same objects shown in the left panel. The solid red line represents the least-squares fit to all the objects. The dashed blue line corresponds to the fit excluding the innermost object, Sh~2-61, that shows a much higher N/O ratio.}
\label{Ngrads}
\end{center}
\end{figure}

In Fig.~\ref{Ngrads} we show the spatial distribution of N abundances of the very low-ionization {\hii} regions of our sample. In these objects we can assume that  N/H $\approx$ N$^+$/H$^+$. Therefore, an  ICF is not necessary for estimating the N/H ratio \cite[see][]{estebangarciarojas18}. The least-squares linear fit to the $R_{\rm G}$ of the objects and their N abundance gives the following radial N  abundance gradient: 

\begin{equation} \label{eq:4} 12 + \log(\mathrm{N/H}) = 8.21(\pm 0.09) - 0.059(\pm 0.009) R_\mathrm{G}. \end{equation}

As in the case of the O/H gradient, the mean difference of the N abundance of the {\hii} regions and the linear fit at their corresponding distance is $\pm$0.06 dex, of the order of the average uncertainty of the abundance determinations, which is about $\pm$0.05 dex. This result indicates that the amount of N in the ISM of the Galactic disc is fairly homogeneous and that any possible local inhomogeneity is not substantially larger than the observational uncertainties. All previous determinations of the radial abundance gradient of N from {\hii} regions have been calculated using ICFs, and give slopes stepper than ours  (Eq.~\ref{eq:4}). For example, \citet{shaveretal83} obtain $-$0.073 $\pm$ 0.013 dex kpc$^{\rm -1}$ and \citet{rudolphetal06} -- using FIR observations -- obtain $-$0.085 $\pm$ 0.020 dex kpc$^{\rm -1}$. 

In the right panel of Fig.~\ref{Ngrads} we show the radial distribution of log(N/O) as a function of $R_{\rm G}$ for the same objects. The least-squares linear fit to the data (represented by the red solid line) is:

\begin{equation} \label{eq:5} \log(\mathrm{N/O}) = -0.63(\pm 0.17) - 0.012(\pm 0.018) R_\mathrm{G}, \end{equation}

\noindent which slope is rather small and can even be considered flat within the uncertainties. As it can be seen, the slope of the N/O gradient depends strongly on  the innermost object -- Sh~2-61 -- that shows the highest N/O ratio. Recalculating the fit excluding that object the slope becomes virtually flat (blue dashed line in Fig.~\ref{Ngrads}).

\begin{equation} \label{eq:6} \log(\mathrm{N/O}) = -0.79(\pm 0.22) + 0.002(\pm 0.021) R_\mathrm{G}. \end{equation}

The flatness of the radial distribution of the N/O ratio between 8 and 16 kpc, permits to assume a constant mean value of log(N/O) = $-$0.77 $\pm$ 0.04 for that range of $R_{\rm G}$. 
All previous determinations of the radial N/O gradient from {\hii} region spectra have been obtained assuming an ICF scheme to determine the N abundance. Using optical data and an ICF(N$^+$), 
\citet{shaveretal83} obtained a rather flat slope of $-$0.006 dex kpc$^{\rm -1}$, in good agreement with our results. Other determinations of the N/O gradient use FIR spectra, are based on an 
assumed ICF(N$^{2+}$) because 
{\fniii} CELs are the only N lines observable in the FIR range. With these kinds of observations, \citet{simpsonetal95} obtained a gradient of log(N/O) with a slope of $-$0.04 $\pm$ 0.01 dex kpc$^{\rm -1}$. However, 
they find a significantly better result using a step fit with two levels, an inner constant value of log(N/O) = $-$0.49 for objects located at $R_{\rm G}$ $<$ 6 kpc and and external one of log(N/O) = $-$0.74 for 6 kpc $<$ $R_{\rm G}$ $<$ 11 kpc. \citet{rudolphetal06}, also based on FIR spectra and assuming an ICF(N$^{2+}$) from 
photoionisation models, obtained similar results. The results based on FIR observations indicate an enhancement of the N/O ratio in the inner part of the Galactic disc but no strong evidence for an overall linear 
radial gradient of this quantity. The almost flat N/O gradient we see in Fig.~\ref{Ngrads} indicates that the bulk of the N is not formed by standard secondary processes at least in most part of the Galactic disc. Further deep optical spectra of low-ionization {\hii} regions located at  $R_{\rm G}$ $<$ 8 kpc are necessary to confirm if other {\hii} regions apart from Sh~2-61 show the enhancement of N/O that FIR data suggest.

\section{The C/H and C/O gradients}

\begin{figure} 
\begin{center}
\includegraphics[angle=0,width=6.0cm]{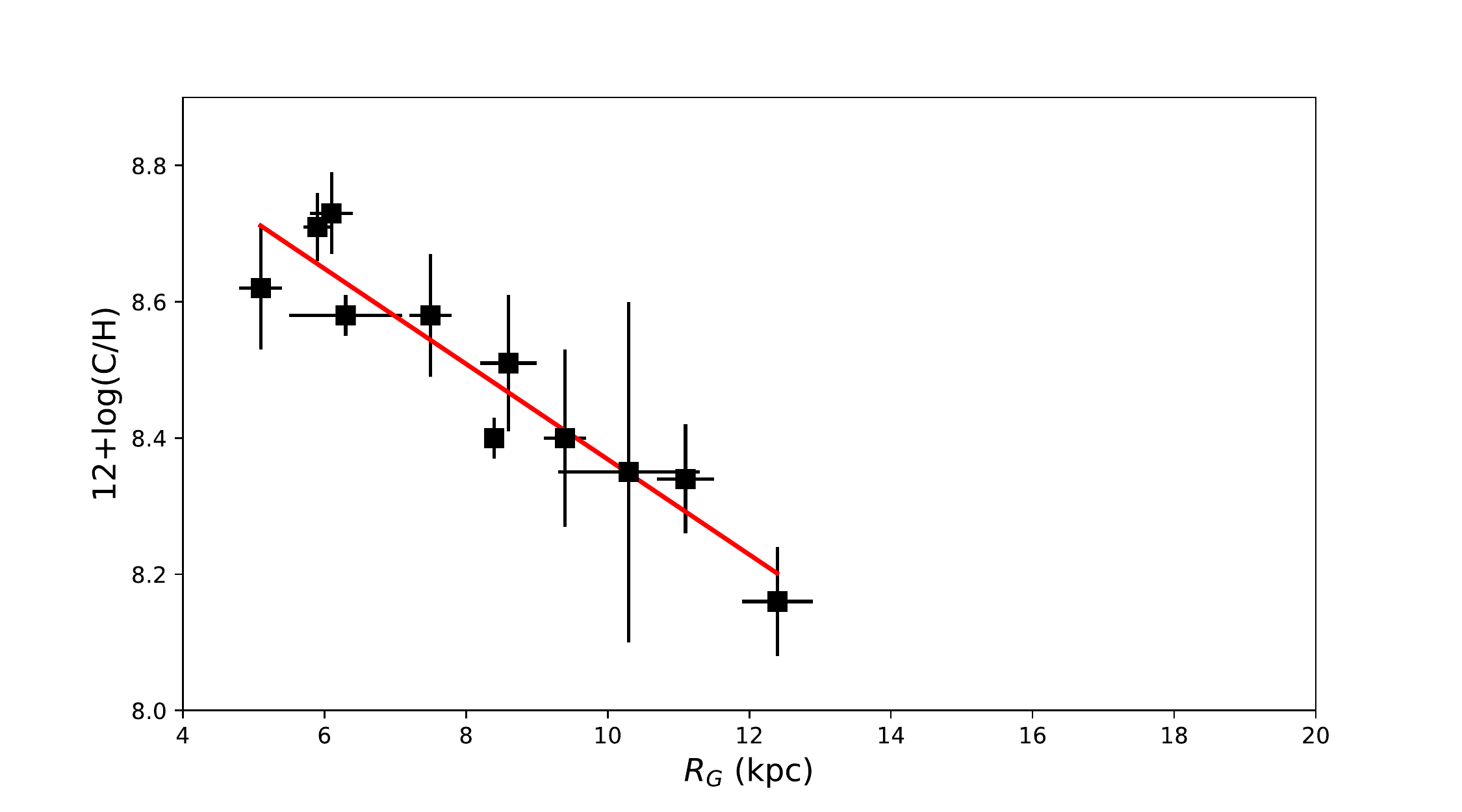}
\hspace*{0.cm}
\includegraphics[width=6.0cm]{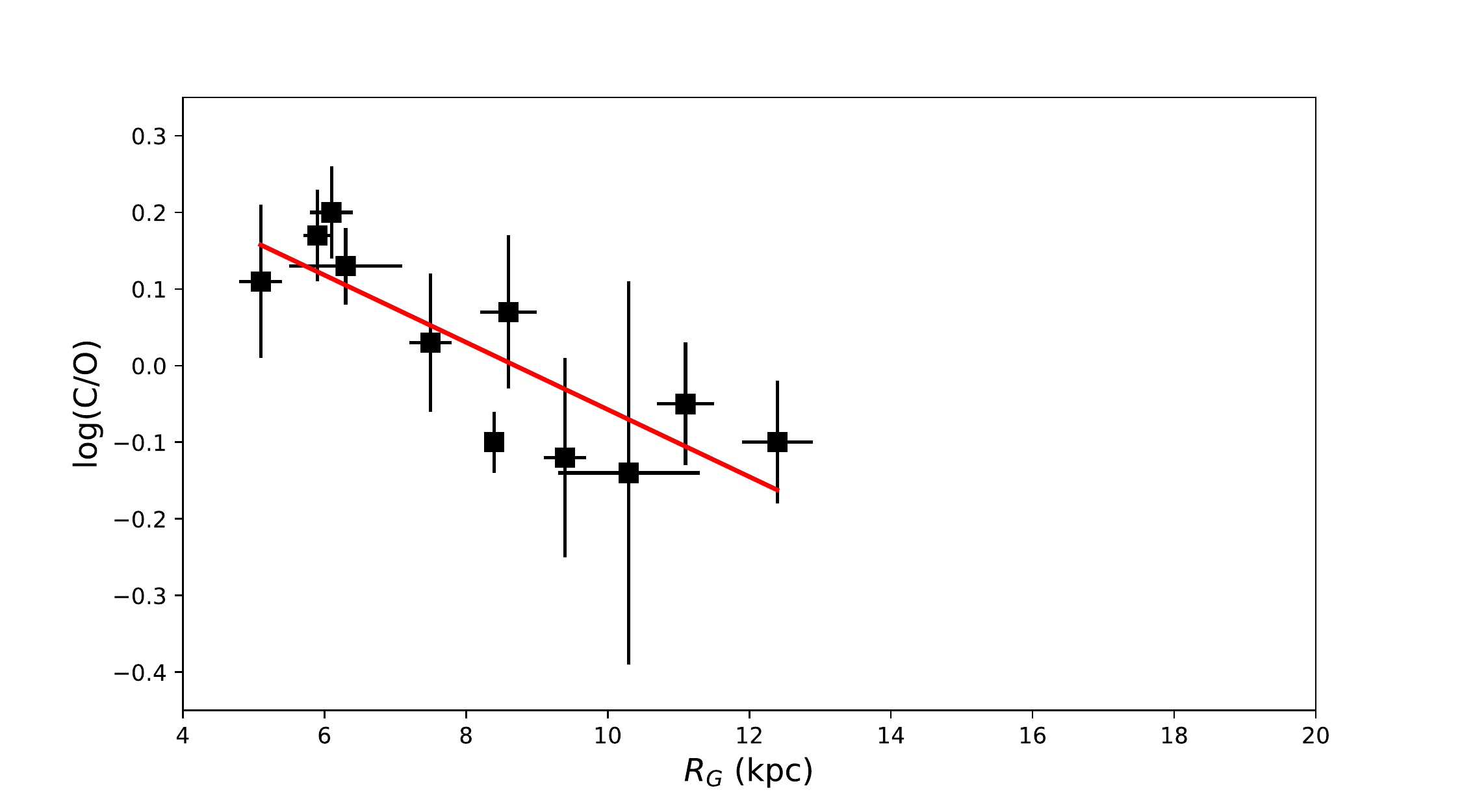}
\caption{ {\it Left:} Radial distribution of the C abundance as a function of $R_{\rm G}$ for our sample of Galactic {\hii} regions. The solid red line represents the least-squares fit to all objects. 
{\it Right:}  Radial distribution of log(C/O) as a function of $R_{\rm G}$ for our Galactic {\hii} region where C and O abundances have been derived from RLs. The solid red line represents the least-squares fit. }
\label{Cgrads}
\end{center}
\end{figure}

\begin{figure} 
\begin{center}
\includegraphics[angle=0,width=6.75cm]{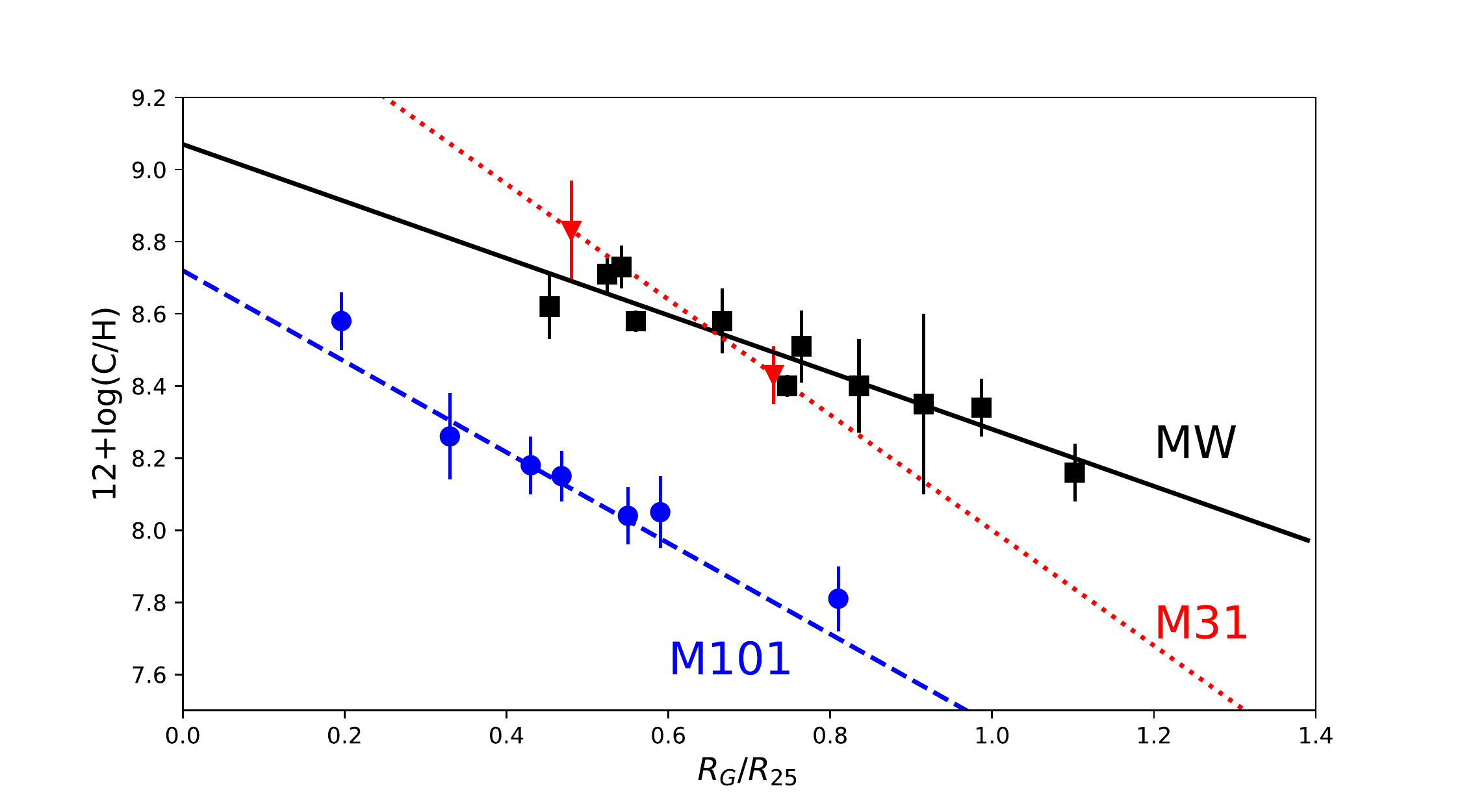}
\hspace*{0.cm}
\includegraphics[width=5.25cm]{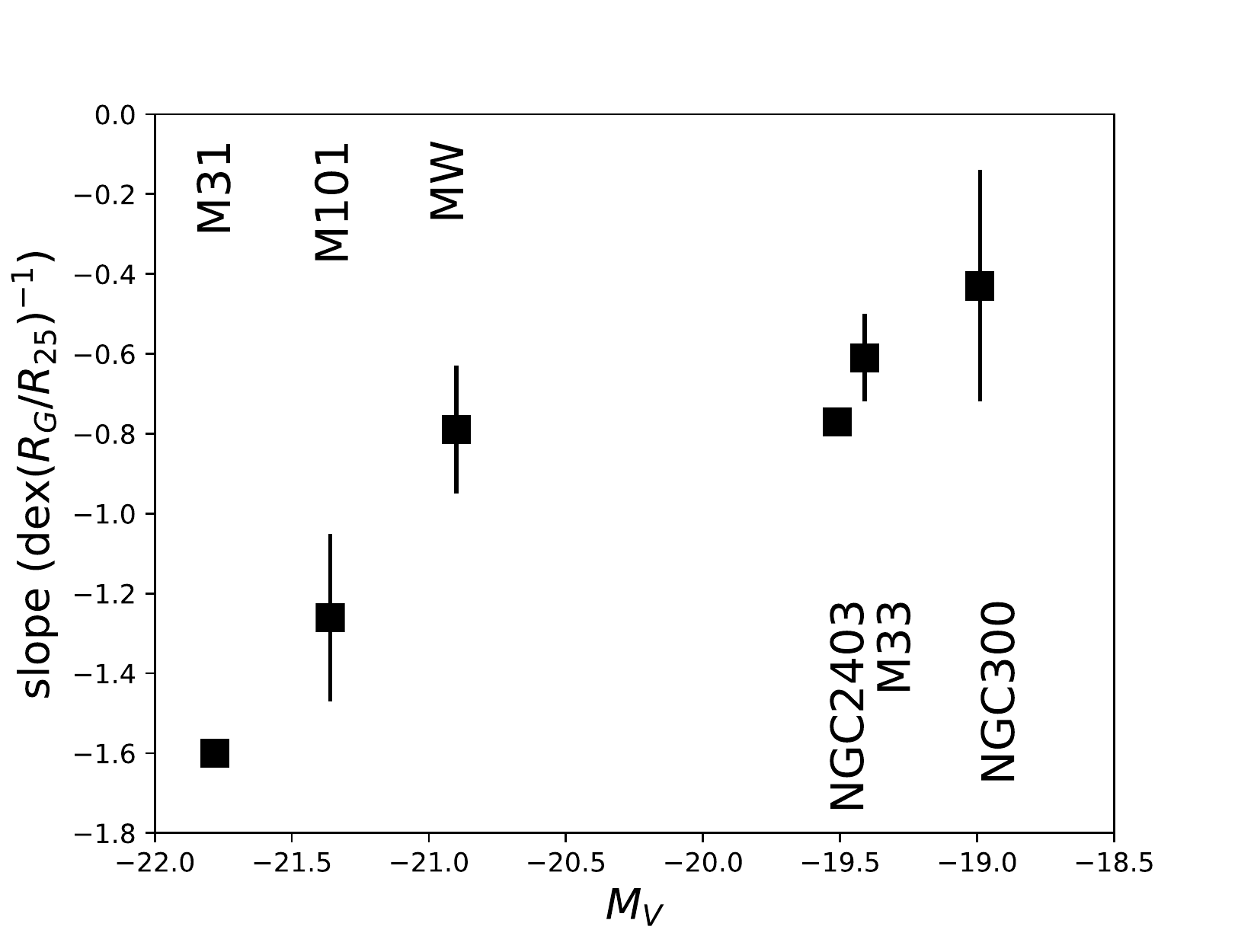}
\caption{ {\it Left:} C radial abundance gradients normalized to galactic isophotal radius $R_{\rm 25}$ obtained from {\hii} region spectra of the Milky Way (black squares, this work) and M101 and M31 (blue circles and red triangles, respectively, data taken from Esteban et al., in preparation). Lines represent the least-squares linear fits for the Milky Way (black solid line), M101 (blue dashed line) and M31 (red dotted line). 
{\it Right:}  Slope of C radial abundance gradients $vs.$ absolute magnitude, $M_\mathrm{V}$ for several spiral galaxies (see text for details).}
\label{Cgalaxies}
\end{center}
\end{figure}

The C/H and C/O gradients of the Milky Way determined from {\hii} region spectra were studied by \cite{estebanetal05, estebanetal13}. In this paper we show a revision of those results including determinations of C/H ratios for two additional objects: Sh~2-100 and Sh~2-152, which data are presented in \cite{estebanetal17} and \cite{estebangarciarojas18}, respectively. The C abundances have been derived from measurements of the faint RL of {\cii} 4267 \AA\ and the application of an ICF(C$^{2+}$) due to the absence of C$^+$ lines in the optical. We have used the mean value of three different estimations of  
ICF(C$^{2+}$) based on the results of photoionization models: the relation found by \cite{garnettetal99} and the recent ones by \cite{bergetal19} and Medina-Amayo et al. (in preparation). In Fig.~\ref{Cgrads} we show the spatial distribution of the resulting C abundances for the {\hii} regions of our sample. The least-squares linear fit to the data (represented by the red solid line) is:

\begin{equation} \label{eq:7} 12 + \log(\mathrm{C/H}) = 9.07(\pm 0.11) - 0.070(\pm 0.014) R_\mathrm{G}, \end{equation}

which is stepper than the O gradient or even the N one. Again, the dispersion of the individual C/H ratios with respect to the fit is rather small --  0.05 dex -- and of the order of the uncertainties, indicating the absence of substantial chemical inhomogeneities along the Galactic disc. In Fig.~\ref{Cgrads} we show the radial distribution of log(C/O) as a function of $R_{\rm G}$ for those objects where we detect both {\cii} and {\oii} RLs in their spectra. This group includes the {\hii} regions studied in \cite{estebanetal05, estebanetal13} and Sh~2-100. The least-squares linear fit to the data (represented by the red solid line in Fig.~\ref{Cgrads}) is:

\begin{equation} \label{eq:8} \log(\mathrm{C/O}) = 0.38(\pm 0.11) - 0.044(\pm 0.014) R_\mathrm{G}. \end{equation}

\cite{toribiosanciprianoetal17} determined the C/H gradients for two nearby low-mass spiral galaxies, NGC~300 and M33, finding also that the C/H gradients are steeper than those of O/H, leading to negative C/O gradients. On the other hand, Esteban et al. (in preparation) are studying the C and O abundance gradients of M31 and M101, which are nearby spiral galaxies more massive than the Milky Way. In Fig.~\ref{Cgalaxies} we compare the C/H gradients of these three galaxies as a function of the fractional galactocentric distance of the {\hii} regions, $R/R_\mathrm{25}$, inside each galaxy. In Fig.~\ref{Cgalaxies} we also represent the slope of the C/H gradient of several spiral galaxies with respect to their $M_\mathrm{V}$ including the data of \cite{toribiosanciprianoetal17} for NGC~300 and M33; of \cite{estebanetal09} for NGC~2403, of Esteban et al. (in preparation) for M101 and M31 and of this work for the Milky Way. In the figure we can see that both quantities are clearly correlated, a result that was firstly reported by \cite{toribiosanciprianoetal17}. There are two galaxies without errorbars in their C/H slope represented in the right panel of Fig.~\ref{Cgalaxies}. The gradient of M31 is based only on two observational points, so we cannot assign a formal error. In the case of NGC~2403, the individual determinations of C/H of the {\hii} regions have large errors and the estimation of the slope is rather uncertain. In the light of the results by \cite{carigietal05}, \cite{toribiosanciprianoetal17} propose that the correlation between the slope of the C/H gradient and $M_\mathrm{V}$ may be the product of metallicity-dependent yields of C. At high metallicities, 12+log(O/H) $\geq$ 8.5, the main contributors to C should be massive stars. However, at intermediate metallicities,  8.1 $\leq$ 12+log(O/H) $\leq$ 8.5, the main contributors to C are low metallicity low-mass stars. On the other hand, the behaviour of the C/O radial abundances gradients are consistent with an ``inside-out'' formation scenario. In the inner parts of galaxies, the low- and intermediate-mass stars had enough time to inject C into the ISM. The newborn massive stars formed in more metal-rich environments can further enrich the ISM via their stellar winds. Therefore, this behavior would certainly produce different C/H gradients in galaxies of different luminosities and average metallicities. Further analysis of the C/H and C/O gradients in the Milky Way will be presented in Arellano-C\'ordova et al. (in preparation).

\section{Other elements}

Our deep spectra allow to explore the radial gradients of other elements as He, Ne, S, Cl or Ar. He is the second most abundant element in the Universe but there are rather few works that have tried to determine its gradient in the Milky Way \cite[e.g.][]{peimbertetal78, shaveretal83, fernandezmartinetal17}. Thanks to the depth of our spectra, we can measure a large number of {\hei} lines in most {\hii} regions 
of our sample. The atomic levels of  {\hei} have two possible quantum states, singlet and triplet. Lines from triplet levels are comparatively much more affected by collisional excitation and radiative transfer effects, so singlet lines are more suitable for obtaining precise abundances of the He$^+$/H$^+$ ratio. In the paper authored by M\'endez-Delgado et al. of these proceedings we describe the methodology and our  preliminary determination of the radial He abundance gradient of the Milky Way. A final study will be presented in M\'endez-Delgado et al. (in preparation). 

Some preliminary results on the gradients of Ne, S and Cl abundances -- especially of Cl/H and Cl/O ratios -- can be found in the paper by Arellano-C\'ordova et al. of these proceedings. In a future paper (Arellano-C\'ordova et al., in preparation) we will explore in depth the behavior of Ne, S, Cl and Ar abundance gradients -- as well as their ratios with respect to O -- carrying 
out a detailed analysis of the most appropriate ICF schemes for each element and discussing the implications on the chemical evolution of the Milky Way. 

\acknowledgments This research is funded by the State Research Agency (AEI), Spanish Ministry of Science, Innovation and Universities (MCIU) and the European Regional Development Fund (ERDF) under grant AYA2015-65205-P. JG-R acknowledges support from an Advanced Fellowship from the Severo Ochoa excellence program (SEV-2015-0548). KZA-C acknowledges support from Mexican CONACYT grant 711183. 

\bibliographystyle{aaabib}
\bibliography{Esteban}

\end{document}